\documentclass[runningheads]{llncs}
\usepackage[T1]{fontenc}
\usepackage{graphicx,verbatim}
\usepackage{booktabs} 
\usepackage{multirow}
\usepackage{bbm} 
\usepackage{amssymb}
\usepackage{wrapfig} 
\usepackage{array}
\usepackage{graphicx}
\usepackage{subcaption}
\usepackage{marvosym}
\usepackage[marginal]{footmisc}

\usepackage{bm}
\usepackage{amsmath}
\usepackage{hyperref}
\usepackage{grffile}
\usepackage{pifont} 
\usepackage{mathtools}
\usepackage{tabularx}
\usepackage{float}
\usepackage{booktabs} 
\usepackage{pifont} 
\usepackage{color}

\usepackage{makecell}
\newcommand{\ourmethod}{{TumorGen}}

\begin{document}
%
\title{\ourmethod: Boundary-Aware Tumor-Mask Synthesis with Rectified Flow Matching}
%
\author{
  Shengyuan Liu\inst{1}$^*$ \and
  Wenting Chen\inst{2}\thanks{Equal contributions.} \and
  Boyun Zheng\inst{1} \and
  Wentao Pan\inst{1} \and \\
  Xiang Li\inst{3}$^\dagger$ \and
  Yixuan Yuan\inst{1}\thanks{Corresponding author: Xiang Li
(xli60@mgh.harvard.edu),  Yixuan Yuan (yxyuan@ee.cuhk.edu.hk)}
}

\institute{
  The Chinese University of Hong Kong \and
    City University of Hong Kong \and 
  Massachusetts General Hospital and Harvard Medical School
}



\maketitle              
\begin{abstract}
Tumor data synthesis offers a promising solution to the shortage of annotated medical datasets. However, current approaches either limit tumor diversity by using predefined masks or employ computationally expensive two-stage processes with multiple denoising steps, causing computational inefficiency. Additionally, these methods typically rely on binary masks that fail to capture the gradual transitions characteristic of tumor boundaries. We present \ourmethod, a novel Boundary-Aware Tumor-Mask Synthesis with Rectified Flow Matching for efficient 3D tumor synthesis with three key components: a Boundary-Aware Pseudo Mask Generation module that replaces strict binary masks with flexible bounding boxes; a Spatial-Constraint Vector Field Estimator that simultaneously synthesizes tumor latents and masks using rectified flow matching to ensure computational efficiency; and a VAE-guided mask refiner that enhances boundary realism. \ourmethod~significantly improves computational efficiency by requiring fewer sampling steps while maintaining pathological accuracy through coarse and fine-grained spatial constraints. Experimental results demonstrate \ourmethod's superior performance over existing tumor synthesis methods in both efficiency and realism, offering a valuable contribution to AI-driven cancer diagnostics.


\end{abstract}

\section{Introduction}

AI-driven tumor diagnostic systems \cite{ai3,natmed,nature2} are revolutionizing early cancer screening and detection, offering unprecedented precision, efficiency, and accessibility in identifying malignancies at their most treatable stages.
While these computational approaches show promising results, their performance heavily relies on large-scale, high-quality annotated tumor data for training \cite{natcancer4,nature3}. However, obtaining such annotations requires extensive manual effort from medical professionals, making it both time-consuming and cost-prohibitive \cite{limit2,limit1}. A potential solution is to synthesize tumor data, offering a scalable approach to expand training datasets without the burden of manual annotation.


\begin{figure}[t]
  \centering
   \includegraphics[width=0.85\linewidth]{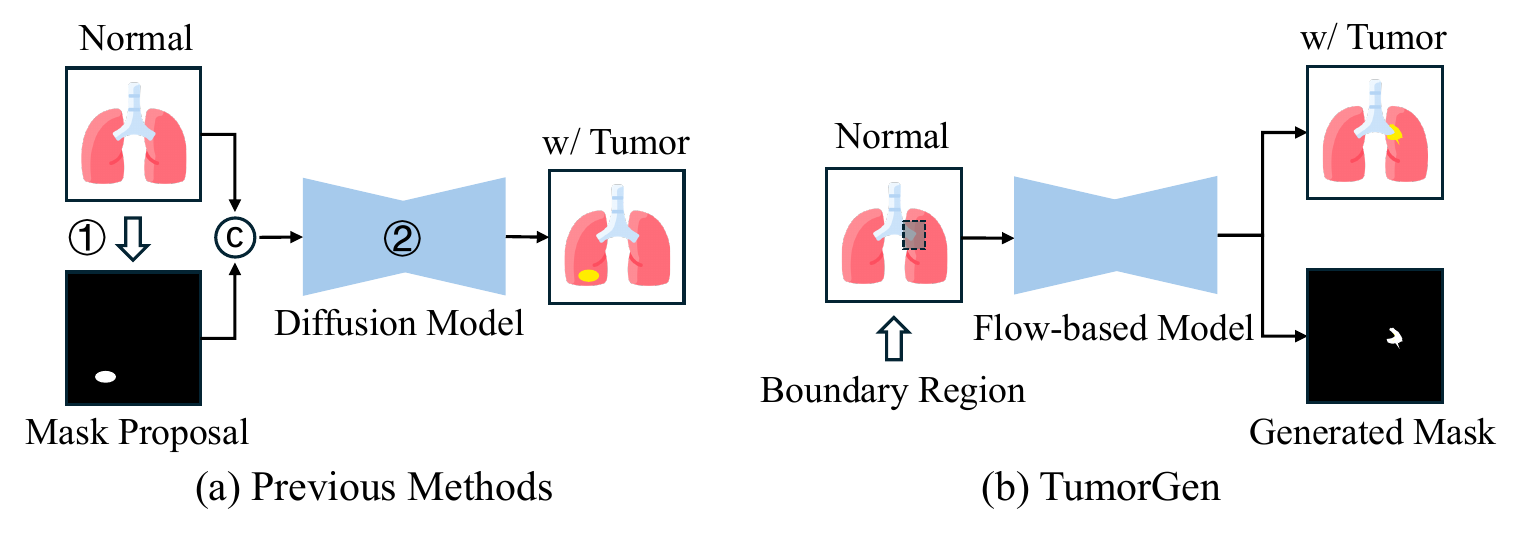}
   \caption{(a) Previous two-stage tumor synthesis methods need to specify the pixel-wise mask. (b) \ourmethod~can synthesize both the tumor image and the corresponding mask with only a rough bounding box.} 
   \label{fig:fig1}
\end{figure}

Current tumor synthesis methods~\cite{DiffTumor,maisi,labelfree,kim2025tumor,peng2024optimizing,Freetumor,zhang2023unsupervised} can be broadly categorized into two groups. The first group utilizes the pre-defined tumor masks as direct input for synthesis~\cite{DiffTumor,maisi,labelfree,peng2024optimizing,Freetumor}, which may limit the diversity of the tumor shape. The second group employs a two-stage approach, where a tumor mask is initially generated followed by tumor synthesis~\cite{kim2025tumor,zhang2023unsupervised}, as shown in Fig.~\ref{fig:fig1} (a). These methods first generate tumor masks by identifying potential tumor locations and creating shapes based on these proposals, then generate tumors according to these masks. The first challenge is that while this two-stage strategy enables more diverse tumor morphologies, its sequential nature introduces significant computational overhead. Additionally, these methods commonly employ diffusion models \cite{DDPM,DDIM} for tumor generation, which require multiple denoising steps (50 to 1,000 steps) during the inference process. Thus, there is a pressing need for more efficient tumor synthesis approaches.

Another challenge is that current methods leverage binary masks for tumor synthesis, ignoring the characteristics of tumor boundaries. These masks can be generated either through rule-based approaches~\cite{DiffTumor,maisi,labelfree,Freetumor} or data-driven methods~\cite{kim2025tumor,zhang2023unsupervised}. Rule-based methods typically generate masks using morphological operations on ellipsoidal primitives, which may not accurately represent the diverse shapes and sizes of real tumors. While data-driven approaches learn to generate masks from real tumor data, they often produce strict boundaries that fail to capture the inherent characteristics of tumor regions. Unlike natural images with clear object boundaries, medical pathologies such as tumors exhibit soft, gradual transitions between diseased and healthy tissues~\cite{softboundary,polyp-Gen}. The binary nature of current mask generation methods overlooks these soft boundaries, potentially compromising the fidelity of synthesized tumors. Therefore, incorporating boundary characteristics into tumor mask generation is crucial for improving the quality of tumor synthesis.

To address these challenges, we introduce a Rectified Flow-based Boundary-Aware Tumor-Mask Synthesis framework, named \textbf{\ourmethod}. Our framework consists of three key components: a \textit{Boundary-Aware Pseudo Mask Generation module (BA-PMG)}, a \textit{Spatial-Constraint Vector Field Estimator (SC-VFE)}, and a \textit{VAE-guided mask refiner (VMR)}. The framework efficiently synthesizes 3D tumors using only a bounding box to indicate potential tumor regions. To consider the \textbf{characteristics of tumor boundary}, the BA-PMG module automatically generates a bounding box for potential tumor regions, replacing the conventional strict binary mask approach and allowing for more flexible tumor generation. In order to enhance \textbf{computational efficiency}, our SC-VFE module simultaneously synthesizes both tumor image latents and their corresponding masks using rectified flow matching, which requires fewer sampling steps by establishing straight mapping between noise and data. Aiming to enhance the realism of generated tumors, the SC-VFE implements two crucial spatial constraints. At a coarse level, we compare ground-truth (GT) and synthetic tumors within the bounding box to ensure overall consistency. For fine-grained control, we leverage the generated tumor mask to align GT and synthetic tumors specifically within the tumor region, ensuring detailed pathological accuracy. To achieve precise and realistic tumor boundaries, our VMR module refines the generated tumor mask using hierarchical features from the 3D-VAE decoder. Finally, 3D-VAE decodes synthetic tumor image latents into the final tumor image. Extensive experimental results demonstrate both the effectiveness of each component and the overall superiority of \ourmethod~compared to existing tumor synthesis methods.

\begin{figure}[t]
  \centering
   \includegraphics[width=\linewidth]{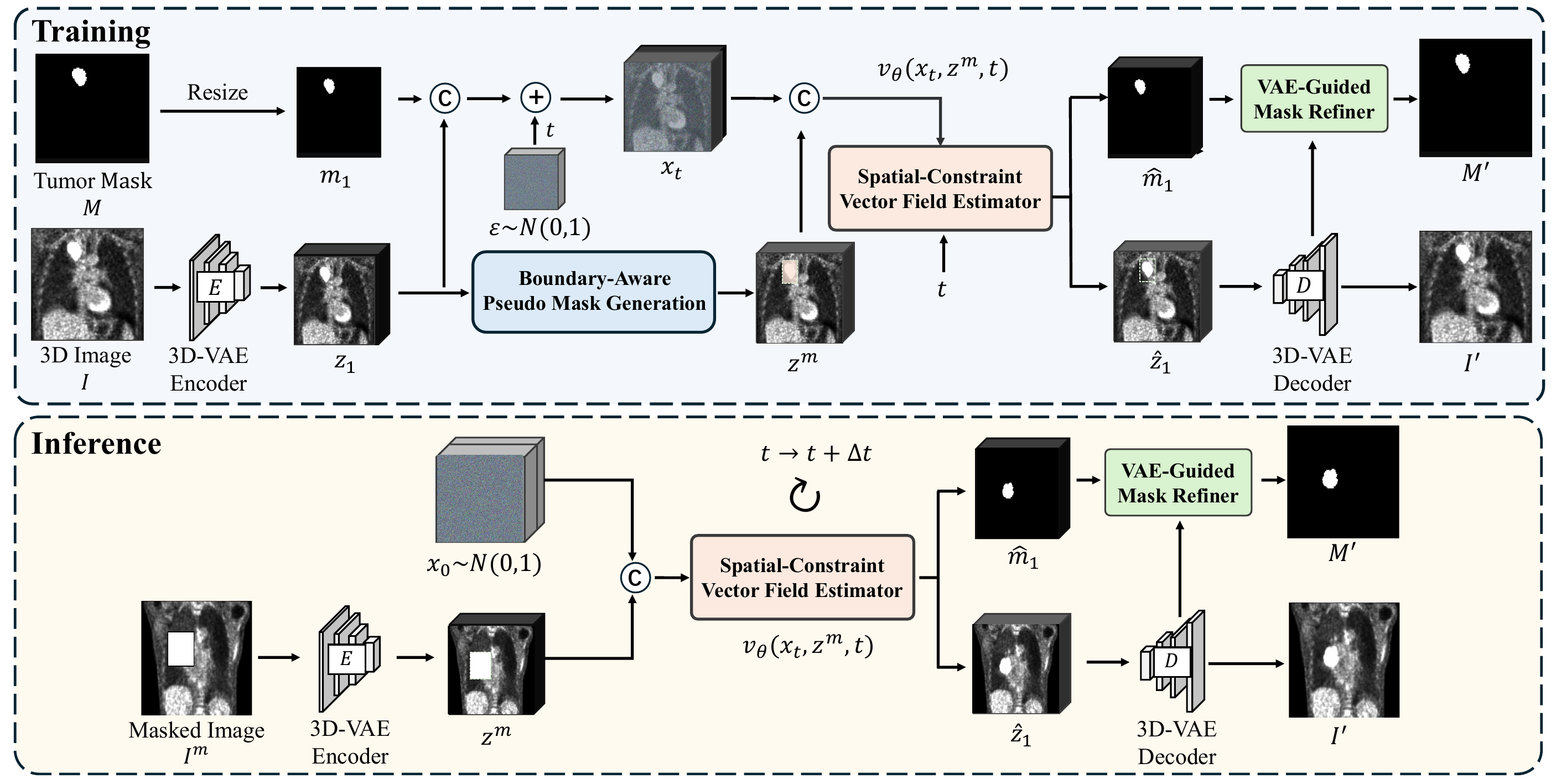}
   \caption{Overview of \textbf{\ourmethod}, a rectified flow-based 3D tumor synthesis framework with a Boundary-Aware Pseudo Mask Generation module (BA-PMG) to automatic tumor bounding box generation, a Spatial-Constraint Vector Field Estimator (SC-VFE) for simultaneous tumor image and mask synthesis, and a VAE-guided mask refiner (VMR) to enhance mask boundary.}
   \label{fig:main}
\end{figure}    


\section{\ourmethod}
\noindent\textbf{Overview.} As shown in Fig.~\ref{fig:main}, given a 3D image $I\in \mathbb{R}^{H \times W \times D}$ and corresponding tumor mask $M\in \{0,1\}^{H \times W \times D}$, we first compress the image using a 3D variational autoencoder (3D-VAE) $E$ to encode the image from the pixel-space into the latent space $z_1 = E(I)\in \mathbb{R}^{\frac{H}{4} \times \frac{W}{4} \times \frac{D}{4} \times 4}$, and downsample the tumor mask $M$ to match the dimensions of $z_1$, obtaining resized tumor mask $m_1$. Subsequently, we employ the Boundary-Aware Pseudo Mask Generation module (BA-PMG) to generate a potential bounding box (bbox) $B$ for the image latents $z_1$, and create a masked image feature $z^m$ by removing the bbox region from $z_1$. We then concatenate $z_1$ and $m_1$ and incorporate initial noise from a normal distribution $\mathcal{N}(0,1)$ into the concatenation using time step $t$ as weight, yielding $x_t = (1-t)\epsilon + t\cdot x_1$. The Spatial-Constraint Vector Field Estimator (SC-VFE) processes the concatenation of $x_t$ and $z^m$ to simultaneously predict the tumor image latents $\hat{z}_1$ and its mask image $\hat{m}_1$. Then, the 3D-VAE decodes $\hat{z}_1$ into a high-quality 3D image $I'$. To enhance tumor boundary precision, the VAE-guided Mask Refiner (VMR) processes the predicted mask image $\hat{m}_1$ while incorporating embeddings from the 3D-VAE of $\hat{z}_1$ to generate the refined mask image $M'$. Finally, 3D-VAE decodes the tumor image latents $\hat{z}_1$ to the final tumor image $I'$.

\subsection{Boundary-Aware Pseudo Mask Generation (BA-PMG)}



Previous tumor synthesis methods
rely on binary masks, failing to capture the gradual transitions between tumor and healthy tissue and causing unrealistic synthetic tumors. Thus, we propose a boundary-aware pseudo mask generation (BA-PMG) approach that automatically generates adaptive bounding boxes around potential tumor regions, enabling more natural tumor synthesis.

Specifically, given a resized tumor mask $m_1$, we first compute a tight bounding box $B'$ encompassing the tumor by identifying the minimum and maximum coordinates of tumor voxels in each dimension. To determine the final bounding box $B$, we define offset distances $d={(d_x,d_x'), (d_y,d_y'), (d_z,d_z')}$ between the planes of $B$ and $B'$, where the pairs represent offsets along the $x$, $y$, and $z$ axes respectively. These offsets are randomly sampled from a uniform distribution within a range proportional to the image dimensions (width $W$, height $H$, and depth $D$) of the 3D image $I$. Specifically, each offset is sampled as $d_i \sim U(0, \alpha D_i)$, where $D_i$ is the respective dimension and $\alpha$ is a scaling factor controlling the maximum extension. The final bounding box $B$ is obtained by expanding $B'$ by these offsets: $B = B' + d$. This adaptive expansion allows the model to consider varying amounts of surrounding tissue context. We then mask the image latents $z_1$ using $B$ to obtain masked latents $z^m = (1-B) \odot z_1$, where $\odot$ denotes element-wise multiplication. 
The masked latents retain information about both the tumor region and its surrounding context, which is crucial for learning realistic boundary transitions. The BA-PMG module enables the synthesis model to learn realistic tumor boundary characteristics during generation, moving beyond the limitations of strict binary foreground annotations. 

\subsection{Spatial-Constraint Vector Field Estimator (SC-VFE)}
Although two-stage tumor synthesis models can generate diverse tumor morphologies,
they suffer from computational inefficiency due to their separate mask generation step. Thus, we introduce a Spatial-Constraint Vector Field Estimator (SC-VFE) that leverages Rectified Flow Matching (RFM) \cite{flowmatching,sd3} to simultaneously generate tumor images and their corresponding masks by establishing straight paths between noise and data. To enhance synthetic tumor realism, we implement both coarse-grained and fine-grained spatial constraints, ensuring overall anatomical consistency while preserving detailed pathological features.

Formally, we define the latent space of tumor image $z$ and its mask image $m$ as a joint variable $x=[z,m]$. We employ RFM to learn a vector field estimator $v_\theta$ that models a reversible mapping from an initial normal distribution $\mathcal{N}(0, 1)$ to the target joint distribution $x \sim p_1(z,m)$ along straight paths. Then, we concatenate tumor image latents $z_1$ (encoded by a 3D VAE encoder) with resized mask image $m_1$. For the forward process over time $t$, we perform element-wise addition between the concatenation output and initial noise sampled from a normal distribution $\mathcal{N}(0,1)$, yielding $x_t = (1-t)\epsilon + t\cdot x_1$, where $x_1 = [z_1, m_1]$ is the concatenation of the tumor image latents $z_1$ and its corresponding mask $m_1$, and $\epsilon \sim \mathcal{N}(0,1)$ denotes sampled random noise.
With $x_t$ and masked image latents $z^m$, vector field estimator $v_{\theta}$ outputs the estimated vector field $v_\theta^z, v_\theta^m$ for synthetic tumor image latents $\hat{z}_1$ and mask images $\hat{m}_1$,
$v_{\theta} \left(x_t, z^m,t\right)=[v_\theta^z, v_\theta^m]$.
Subsequently, the synthetic tumor image latents $\hat{z}_1$ and its mask image $\hat{m}_1$ are computed by:
\begin{equation}
\hat{z}_1 = z_t + (1-t)\cdot v_\theta^z,~\hat{m}_1 = m_t + (1-t)\cdot v_\theta^m, 
\end{equation}
\begin{equation}
z_t = (1-t)\epsilon + t\cdot z_1,~m_t = (1-t)\epsilon + t\cdot m_1, 
\end{equation}
To optimize the vector field estimator, we utilize the RFM \cite{sd3} objective function, which is defined as:
\begin{equation}
\mathcal{L}_{rfm}(\theta)=\mathbb{E}_{t,\epsilon\sim\mathcal{N}(0,1),x\sim p_1(z,m)}\left\|v_{\theta} \left(x_t, z^m,t\right)-\left(x_1-\epsilon\right)\right\|^2
\end{equation}
To improve the realism of synthetic tumors, we impose a coarse-grained spatial constraint and a fine-grained spatial constraint on the synthetic tumor image latents, respectively. To preserve the overall consistency, we propose a coarse-grained spatial constraint $\mathcal{L}_{coarse}$, which utilizes the final bounding box generated by BA-PMG module to minimize the difference between the ground-truth (GT) image latents $z_1$ and the synthetic tumor image latents $\hat{z}_1$ inside the bounding box, 
\begin{equation}
\mathcal{L}_{coarse}(\theta)=\mathbb{E}_{t,\epsilon\sim\mathcal{N}(0,1),x\sim p_1(z,m)}\left\|B \odot \hat{z}_1 - B \odot z_1
\right\|^2.
\end{equation}
To preserve the detailed pathological accuracy, we design a fine-grained spatial constraint $\mathcal{L}_{fine}$ to leverage the generated tumor mask to narrow the Structural Similarity (SSIM) distance between GT image latents $z_1$ and synthetic tumor image latents $\hat{z}_1$ inside the tumor mask, 
\begin{equation}
\mathcal{L}_{fine}(\theta)=\mathbb{E}_{t,\epsilon\sim\mathcal{N}(0,1),x\sim p_1(z,m)}[1 - SSIM(\hat{m}_1 \odot \hat{z}_1,{m}_1 \odot z_1
)].
\end{equation}
By adopting the SC-VFE module, we can enable the simultaneous synthesis of tumor and mask images and preserve the realism of the synthetic tumor at both coarse and fine grained levels, thereby eliminating the need for manually specifying pixel-wise tumor mask.

\subsection{VAE-Guided Mask Refiner (VMR)}
Since flow matching occurs in latent space, the generated tumor masks have poor resolution and cannot properly align with the VAE decoder's output images.
To generate precise tumor mask images, we propose a VAE-Guided Mask Refiner (VMR) that leverages the spatial information from the VAE model to refine tumor mask boundaries.

Specifically, a 3D VAE decoder transforms synthetic tumor image latents $\hat{z}_1$ into a tumor image $I'$ and extracts corresponding features $f=\left\{f_1, f_2, f_3\right\}$ from three different decoder layers. These hierarchical features $f$ are processed through convolutional layers to obtain $f'=\left\{f_1', f_2', f_3'\right\}$. The VMR module $f_\theta$ then takes the generated tumor mask $\hat{m}_1$ as input and performs element-wise addition between its layer features $q=\left\{q_1, q_2, q_3\right\}$ and the corresponding features in $f'=\left\{f_1', f_2', f_3'\right\}$ to produce the refined tumor mask $M'$.
To optimize the refined tumor mask, we compute the reconstruction loss between the ground truth and refined mask image:
\begin{equation}
\mathcal{L}_{rec}(\theta)=\mathbb{E}_{t,\epsilon\sim\mathcal{N}(0,1),x\sim p_1(z,m)}\left\| f_\theta(\hat{m}_1) - M \right\|^2,
\end{equation}
where $f_\theta(\hat{m}_1)$ represents the refined tumor mask $M'$, and $M$ is the ground truth tumor mask. This loss function ensures high-resolution mask generation that closely aligns with the image content by minimizing the difference between the generated and ground truth masks.

Finally, the overall objective function for the proposed method integrates four components: the rectified flow matching objective function, coarse-grained spatial constraint, fine-grained spatial constraint, and reconstruction loss for the refined mask image. These components are combined as:
\begin{equation}
\mathcal{L} = \mathcal{L}_{rfm} + \lambda_1 \mathcal{L}_{coarse} + \lambda_2 \mathcal{L}_{fine} + \lambda_3 \mathcal{L}_{rec},
\end{equation}
where $\lambda_1$, $\lambda_2$, and $\lambda_3$ are weighting coefficients.

\section{Experiments}

\subsection{Experimental Setting}

\noindent\textbf{Datasets.} 
The experiments use two public available Positron Emission Tomography (PET) datasets: Hecktor-2021~\cite{hecktor2021} with 224 PET scans (200 training, 24 testing) and AutoPET~\cite{AutoPET} with 168 whole-body PET scans from lung cancer patients (153 training, 15 testing). For AutoPET, TotalSegmentator~\cite{totalsegmentator} is used to segment lung regions, which were then cropped to $200\times200\times100$ and resized to $128\times128\times64$ for training. All scans include both PET images and corresponding tumor segmentation masks.

\noindent\textbf{Implementation details and metrics.} 
We use a pre-trained 3D KL-VAE~\cite{maisi} as 3D-VAE. For spatial-constraint vector field estimation, we implement a 3D U-Net architecture. 
For evaluation, image quality is evaluated using Fréchet Inception Distance (FID)~\cite{FID}. We calculate the FID on the middle slices on axial, coronal, and sagittal planes of the generated 3D images following \cite{3d-stylegan}. Moreover, to evaluate the alignment quality of synthesized image-mask pairs by TumorGen, we use a pre-trained nnU-Net \cite{nnUNet} on the synthesized images to obtain predicted masks, and calculate both Dice Similarity Coefficient (DSC) and Normalized Surface Distance (NSD) with the generated masks.


\begin{table}[t]
\fontsize{8pt}{10pt}\selectfont
\setlength{\tabcolsep}{0.9mm}
\centering
\caption{Quantitative evaluation on the Hecktor and AutoPET datasets. A, S, C denotes Axial, Sagittal, and Coronal planes. Avg denotes the average result.}

\label{tab:sota}
\begin{tabular}{ccccccccc}
\toprule
\multirow{2}{*}{\textbf{Method}} & \multicolumn{4}{c}{\textbf{Hecktor}} & \multicolumn{4}{c}{\textbf{AutoPET}} \\
\cmidrule(lr){2-5} \cmidrule(lr){6-9}
& FID(A) & FID(S)  & FID(C) & FID(Avg) & FID(A) & FID(S)  & FID(C) & FID(Avg)  \\
\midrule
HA-GAN \cite{HAGAN} & 137.561 & 103.794 & 114.298 & 118.551 &    130.187 & 152.133 & 128.709 & 137.010 \\
MedFusion \cite{thambawita2022singan} & 144.625 & 125.834 & 87.517 & 119.325 & 138.462 & 129.524& 136.280 & 134.755\\
BLD \cite{blendedDiffusion} & 75.718 & 56.883 & 73.486  & 68.696 & 72.371 & 114.110 & 109.307 & 98.596 \\
SynTumor \cite{labelfree} & 93.024 & 85.035 & 91.042 & 89.701 & 108.671 & 98.704 & 119.638 & 109.004\\
DiffTumor \cite{DiffTumor}  & 80.281 & 52.659 & 67.523 & 66.821 &   82.394 & 96.248 & \textbf{76.195} & 84.946  \\
\midrule
\ourmethod &  \textbf{54.877} & \textbf{39.633} & \textbf{61.593}  &  \textbf{52.035} & \textbf{66.432} & \textbf{66.933} & 83.934 & \textbf{72.433} \\
\bottomrule
\end{tabular}
\end{table}

\begin{figure}[t]
  \centering
   \includegraphics[width=\linewidth]{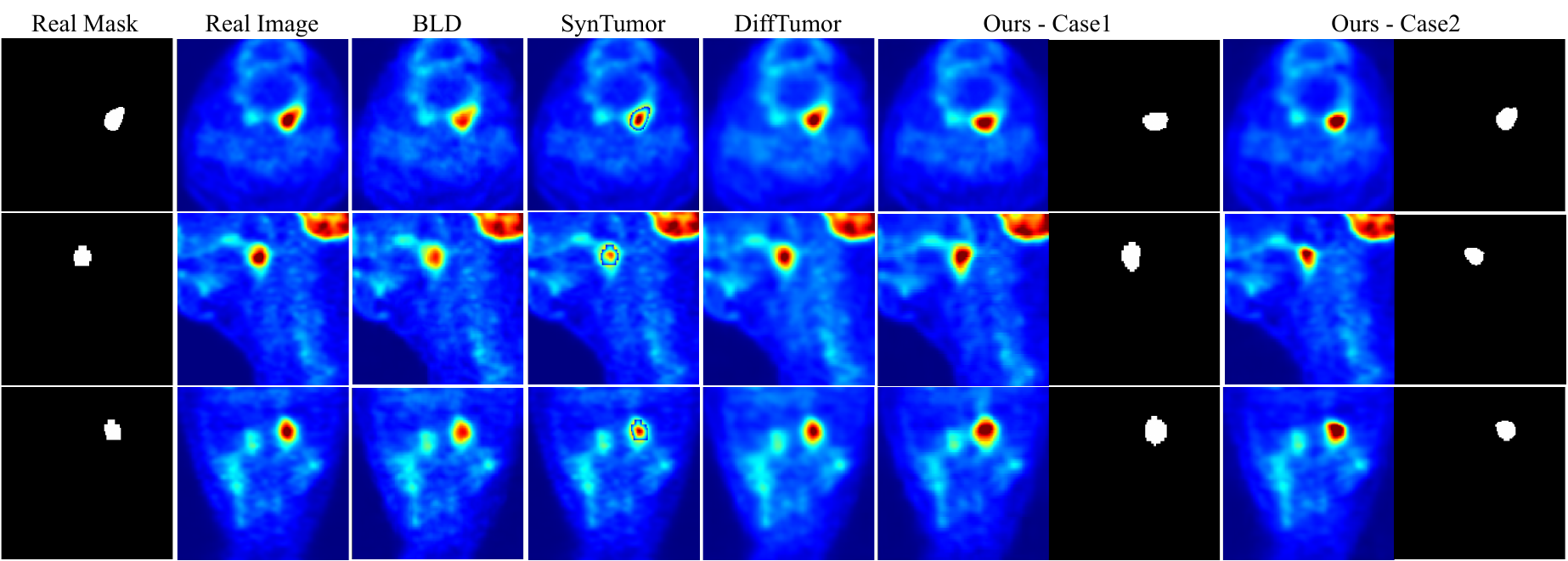}
   \caption{Qualitative comparison among different synthesis methods.}
   \label{fig:visual}
\end{figure}    

\begin{figure}[t]
  \centering
   \includegraphics[width=\linewidth]{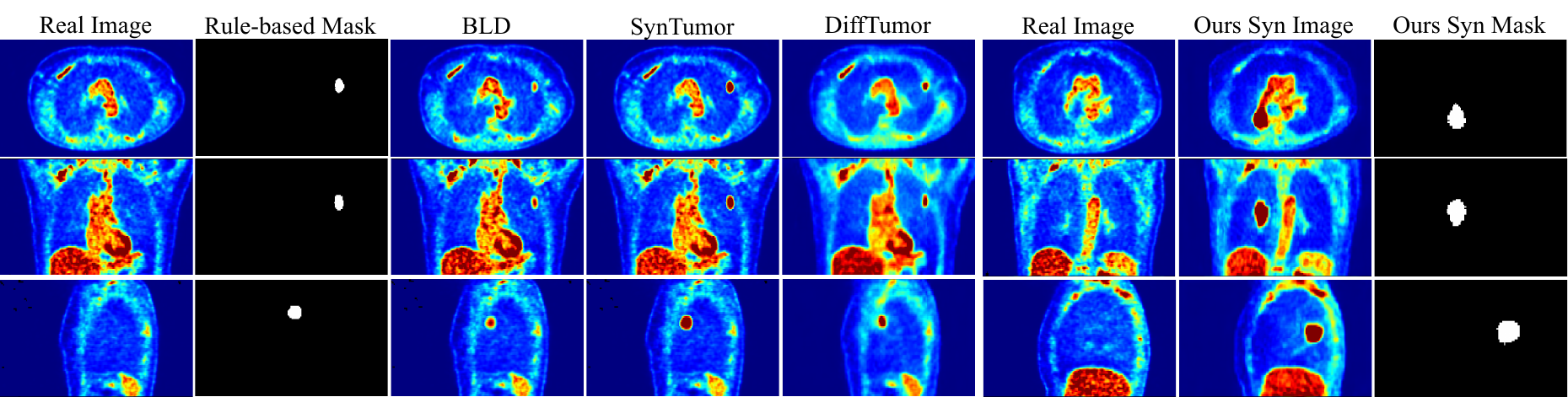}
   \caption{Tumor images generated from healthy images. 
   }
   \label{fig:normal}
\end{figure}    

\subsection{Experimental Results}

\noindent\textbf{Comparison with Previous Methods.} We compare our \ourmethod~with existing tumor synthesis methods, including the conditional generation methods (HA-GAN \cite{HAGAN} and MedFusion \cite{meddiffusion_scireport}), and the two-stage methods that generate tumors by pre-defined masks (SynTumor \cite{labelfree}, BLD \cite{blendedDiffusion}, and DiffTumor \cite{DiffTumor}). In Table~\ref{tab:sota}, our \ourmethod~surpasses the DiffTumor~\cite{DiffTumor} by 14 in average FID on the Hecktor dataset, implying our superiority to existing methods. For qualitative evaluation, Fig.~\ref{fig:visual} shows that \ourmethod~can synthesize realistic tumors with better tumor boundaries and also provide great diversity of the tumor shape.

\noindent\textbf{Ablation Study.} To prove the effectiveness of each module in \ourmethod, we conduct the ablation study on the Hecktor dataset~\cite{hecktor2021}. We evaluate the synthetic tumor image quality using FID score, and Table~\ref{tab:ablation} shows that ablating each component leads to performance decreases, suggesting the effectiveness of each component. Moreover, we evaluate the alignment of generated image-mask pairs using DSC and NSD metrics. The results demonstrate that our generated tumor masks maintain high consistency with the corresponding images, and the introduction of VMR significantly enhances this alignment.

\noindent\textbf{Efficiency Analysis.} We compute the inference time (seconds) for DiffTumor~\cite{DiffTumor} with 50-step DDPM \cite{DDPM} and DDIM \cite{DDIM} samplers, and compare with our method using both 10 and 50 sampling steps with RFM. Table~\ref{tab:sampler} shows that our method generates each sample in 0.218s using a 10-step RFM, while achieving an FID of 55.862, highlighting the computational efficiency of \ourmethod.

\noindent\textbf{Generalizability to Healthy Subjects.} We evaluate the generalization capability of \ourmethod~in the healthy subjects. As shown in Fig.~\ref{fig:normal}, \ourmethod~can synthesize the realistic tumors in the lung region across three planes, suggesting its strong capability in generalizing in PET imaging of healthy patients.


\begin{table}[t]
    \centering

    \begin{minipage}{0.51\textwidth}
        \centering
        \footnotesize
        \setlength\tabcolsep{1.1pt}
        \caption{Ablation study on Hecktor dataset.}
        \label{tab:ablation}
        \scalebox{0.91}{
        \begin{tabular}{ccccccc}
            \toprule
          \multirow{2}{*}{\textbf{\makecell{BA-PMG}}} & \multicolumn{2}{c}{\textbf{SC-VFE}} & \multirow{2}{*}{\textbf{VMR}} & \multicolumn{3}{c}{Metrics} \\  \cmidrule(lr){2-3}\cmidrule(lr){5-7} 
          
             & Coarse & Fine &  & FID & DSC & NSD \\ \midrule
             \ding{55} & \ding{51}  & \ding{51} & \ding{51} & 66.263 & 0.677 & 0.711 \\
          \ding{51}  &    \ding{55}      & \ding{51}  & \ding{51} & 55.416 & 0.660 & 0.673 \\
          \ding{51} & \ding{51} &\ding{55} & \ding{51} & 63.611 & 0.680 & 0.713\\
           \ding{51} & \ding{51} & \ding{51}  & \ding{55} & 54.279 & 0.614 & 0.577 \\
            \midrule
           \ding{51} & \ding{51} & \ding{51}  & \ding{51} & \textbf{52.035} & \textbf{0.694} & \textbf{0.741} \\
            \bottomrule
        \end{tabular}}
    \end{minipage}
    \hfill
    \begin{minipage}{0.43\textwidth}
        \centering
        \footnotesize 
        \caption{Comparison of inference time.}
        \label{tab:sampler}
        \setlength\tabcolsep{1.1pt}
        \scalebox{0.91}{
        \begin{tabular}{cccc}
            \toprule
            Methods & \makecell{Sampling}  & FID & \makecell{Inference\\Time (s)}\\
            \midrule
            DiffTumor & DDPM-50 & 66.821 & 1.226 \\
            DiffTumor & DDIM-50 &  68.306 & 1.133 \\
            \ourmethod      & RFM-50   & \textbf{52.035} & 1.061\\
            \ourmethod      & RFM-10   & 55.862 & \textbf{0.218} \\ 
            \bottomrule
        \end{tabular}}
    \end{minipage}
\end{table}

\section{Conclusions}
\ourmethod~addresses key limitations in tumor synthesis through three innovative components: BA-PMG, SC-VFE, and VMR. Our experiments on two PET datasets show superior performance over existing methods with significantly improved FID scores and strong segmentation metrics. TumorGen achieves greater computational efficiency while maintaining high-quality outputs and demonstrates effective generalization to healthy subjects. 
\bibliographystyle{splncs04}
\bibliography{refs}
\end{document}